\documentclass[prb,aps,twocolumn,superscriptaddress,showpacs]{revtex4}

\usepackage{graphicx}

\begin{document}

\title{Neutron and X-ray diffraction study of cubic [111] field cooled
Pb(Mg$_{1/3}$Nb$_{2/3}$)O$_{3}$}

\author{C. Stock}
\affiliation{Department of Physics and Astronomy, Johns Hopkins University,
Maryland, 21218}

\author{Guangyong Xu}
\affiliation{Condensed Matter Physics and Materials Science Department, Brookhaven National Laboratory, Upton, New
York, 11973}

\author{P. M. Gehring}
\affiliation{NIST Center for Neutron Research, National Institute of
Standards and Technology, Gaithersburg, Maryland, 20899}

\author{H. Luo}
\affiliation{Shanghai Institute of Ceramics, Chinese Academy of Sciences,
Shanghai, China, 201800}

\author{X. Zhao}
\affiliation{Shanghai Institute of Ceramics, Chinese Academy of Sciences,
Shanghai, China, 201800}

\author{H. Cao}
\affiliation{Department of Materials Science and Engineering, Virginia Tech.,
Blacksburg, Virginia, 24061}

\author{J.F. Li}
\affiliation{Department of Materials Science and Engineering, Virginia Tech.,
Blacksburg, Virginia, 24061}

\author{D. Viehland}
\affiliation{Department of Materials Science and Engineering, Virginia Tech.,
Blacksburg, Virginia, 24061}

\author{G. Shirane}
\affiliation{Physics Department, Brookhaven National Laboratory, Upton, New
York, 11973}

\date{\today}

\begin{abstract}

Neutron and x-ray diffraction techniques have been used to study the
competing long and short-range polar order in the relaxor
ferroelectric Pb(Mg$_{1/3}$Nb$_{2/3}$)O$_{3}$ (PMN) under a [111]
applied electric field. Despite reports of a structural transition
from a cubic phase to a rhombohedral phase for fields E $>$ 1.7 kV/cm, we find that the bulk
unit cell remains cubic (within a sensitivity of
90$^{\circ}$-$\alpha$ =0.03$^{\circ}$)for fields up to 8 kV/cm.
Furthermore, we observe a structural transition confined to the near
surface volume or `skin' of the crystal where the cubic cell is 
transformed to a rhombohedral unit cell at T$_{c}$=210 K for E $>$ 4 kV/cm, for which 90$^{\circ}$-$\alpha$=0.08 $\pm$
0.03$^{\circ}$ below 50 K. While the bulk unit cell remains cubic, a suppression of the diffuse
scattering and concomitant enhancement of the Bragg peak intensity
is observed below T$_{c}$=210 K, indicating a more ordered structure
with increasing electric field yet an absence of a long-range
ferroelectric ground state in the bulk.  The electric field strength
has little effect on the diffuse scattering above T$_{c}$, however
below T$_{c}$ the diffuse scattering is reduced in intensity and
adopts an asymmetric lineshape in reciprocal space. The absence of
hysteresis in our neutron measurements (on the bulk) and the
presence of two distinct temperature scales suggests that the ground
state of PMN is not a frozen glassy phase as suggested by some
theories but is better understood in terms of random fields
introduced through the presence of structural disorder.  Based on
these results, we also suggest that PMN represents an extreme
example of the two-length scale problem, and that the presence of a
distinct skin maybe necessary for a relaxor ground state.

\end{abstract}


\maketitle

\section{Introduction}

    The relaxor ferroelectrics have attracted considerable interest due to
their unique dielectric properties and exceptionally large
piezoelectric coefficients.~\cite{Ye98:81,Bokov06:41,Hirota06:75,Park97:82}
Pb(Zn$_{1/3}$Nb$_{2/3}$)O$_{3}$ (PZN) and
Pb(Mg$_{1/3}$Nb$_{2/3}$)O$_{3}$ (PMN) are prototypical relaxors
which, under zero field cooling, show a broad and frequency
dependent peak in the dielectric constant but no well defined
structural transition.   The broad anomaly of the dielectric
response is reflected in the phonon spectra where a soft transverse
optic mode softens and then recovers with decreasing
temperature.~\cite{Wakimoto02:65}  Concurrent with the recovery of
the soft-optic mode, strong diffuse scattering is observed along the
$\langle$110$\rangle$ directions, indicative of short-range
ferroelectric order, which persist to low temperature.  Measurements
using neutron pair distribution function analysis have confirmed
that at low temperatures no long-range polar order is present with a
maximum of $\sim$ 1/3 of the sample having polar
order.~\cite{Jeong05:94}

    Under the application of a strong electric field, a sharp and frequency-independent peak in the dielectric response is observed to remain after the
removal of the electric field at low temperatures indicative of a well defined
structural distortion.~\cite{Ye93:145}  However, NMR
measurements have shown that it is difficult to associate this with
the presence of a long-ranged ferroelectric ground state as two components are
measured in the NMR lineshape and at least half of the crystal remains in an
disordered state.~\cite{Blinc03:91}  Dielectric measurements have found the
presence of a frozen glassy state at low temperatures and have led to theories
of random bond and field effects in analogy to the case of spin
glasses.~\cite{Gvasaliya05:17,Westphal92:68,Pirc04:69,Fish03:67}

    Recent neutron and x-ray scattering studies conducted on PMN and PZN in zero
field have found no long-range structural distortion in the bulk at
low temperatures.  This new phase (referred to as phase-X in the 
original measurements) was first measured and presented
based on neutron diffraction measurements on PZN doped with 8\%
PT.~\cite{Ohwada03:67,Gehring04:70} Instead of long-range 
ferroelectric order, strong anisotropic
diffuse scattering around the Bragg peaks starts near the same
temperature where index of refraction measurements indicate that
local polar regions are formed.~\cite{Burns93:48,Cross87:76} Several
models describing the diffuse scattering in terms of phase-shifted
polar nanoregions and static strain fields have been
proposed.~\cite{Vak04:xx,Hirota02:65,Welberry05:38,You97:79}  We
emphasize that all of these models relate the diffuse scattering
to the presence of polar correlations in the material. Therefore
investigating the effects of an external electric field on the diffuse scattering,
direct information of the local polar correlations can be obtained.

    The diffuse scattering, the lattice dynamics, and the dielectric response
in the relaxors PMN and PZN can be described by two temperature scales.
At a high temperature (T$_{d}$), short-range polar correlations begin to
develop giving rise to strong diffuse scattering measured through
the neutron and x-ray elastic scattering cross sections.  The phonons also broaden
in energy significantly for temperatures below this onset, which is indicative
of very short lifetimes and strong damping.~\cite{Gehring01:63}  Upon cooling below T${c}$ (defined as the temperature where a sharp dielectric peak is observed under an applied field), the phonons begin to sharpen in energy and their frequencies harden while the diffuse scattering persists.
Understanding the response of the diffuse scattering to an electric field as a function
of temperature is central to understanding
the role of T$_{c}$ in the structure of PMN.  We note that in PMN
T$_{c}$=210 K and T$_{d}\sim$690 K.

    Recent electric field studies of pure PZN and PZN-8\% PT have shown
several interesting results.~\cite{Xu05:72} Upon cooling
below T$_{c}$, the diffuse scattering intensity is redistributed
along different $\langle110\rangle$ directions while the integrated
intensity is conserved. Like the induced polarization, the
redistribution of the diffuse scattering persists at low
temperatures after removal of the field, which suggests that there
is a direct connection between it and the formation of ferroelectric
macro-domains under field.  Extensive research investigating the
shape of the unit cell under the application of an electric field
near the morphotropic phase boundary has been presented
elsewhere.~\cite{Cao06:100}  In the case of pure PMN, previous work
has shown a suppression of neutron diffuse scattering intensities occurs in the direction transverse to $\vec{Q}$ across various Bragg peaks with the
application of a [110] field, but no correlation between this effect
and the presence of ferroelectric domains and/or lattice distortions
has been studied.~\cite{Vak98:40}

    In this paper we will show that the electric field reduces the diffuse
scattering in PMN and enhances Bragg peak intensities at low temperatures,
while the unit cell shape remains cubic in the bulk for field
strengths up to 8kV/cm. In contrast to previous experiences with PZN and PZN-8$\%$PT, there
is no field memory effect as the diffuse scattering and Bragg peak
intensities recover their zero-field cooled values at low
temperatures immediately upon removal of the external field.
Low-energy x-ray diffraction measurements, which probe the skin of
the sample, demonstrate that the near-surface region does undergo a
structural transition from a cubic to rhombohedral phase.  The
distortion remains after removal of the field at low-temperatures
and therefore displays the expected memory effect based on
dielectric and polarization measurements. Our results illustrate
that at low temperatures the long-range atomic order, i.e.,
correlations between polar domains, can be enhanced by an external
field. However the intrinsic structural disorder prevents a phase
transition from occurring to a long-range ferroelectric state.

    This paper is divided into two sections.  In the first section we present cold and thermal
neutron data that characterize the diffuse scattering (and hence
the short-range polar correlations) around the $\vec{Q}$=(001) and
$\vec{Q}$=(003) reciprocal lattice positions as a function of
electric field. The second section deals with the possibility of a
structural distortion.  We use neutrons (which probe the entire
crystal volume) to study the shape of the bulk unit cell, and
low-energy x-rays to study the shape of the unit cell near the
surface of the crystal.

\begin{figure}[t]
\includegraphics[width=9cm] {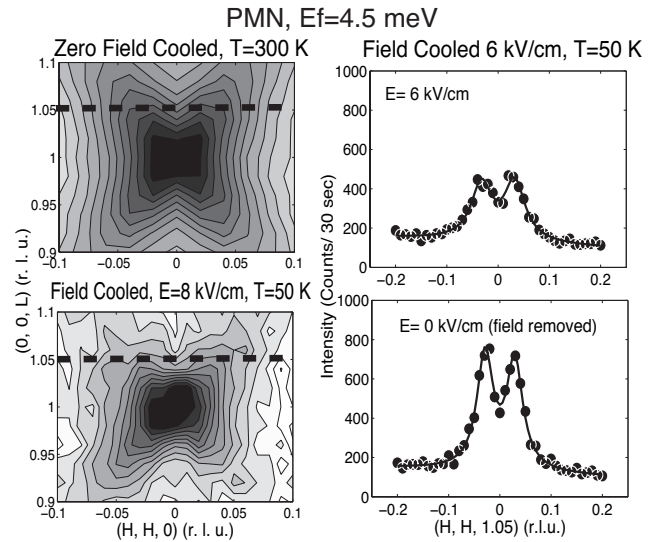}
\caption{\label{diffuse_001} The diffuse scattering in zero field
cooled and field cooled states around the (001) Bragg peak.  The dark regions represent high intensity contours in comparison to lighter regions.  The
dashed lines illustrate the line scans conducted to investigate the
temperature dependence of the intensity.}
\end{figure}

\section{Experiment}

    Neutron scattering experiments were conducted at the BT9 thermal and SPINS
cold triple-axis spectrometers located at the NIST Center for
Neutron Research on a  1 cm $\times$ 1 cm $\times$ 0.1 cm plate
aligned in the (HHL) scattering plane. Sputtered gold electrodes
allowed an electric field to be applied along the [111] direction
and independent polarization measurements have revealed a non-zero
polarization at low temperatures after field cooling, consistent
with previous work.~\cite{Ye93:145,Cao:unpub}  The electric field
was measured during the experiments through the use of two wires
attached in locations different from the two wires driving
the voltage difference across the crystal. The measured voltage was
always the same as that applied at the power supply. Field cooling
sequences were done by first heating the sample to temperatures
between 550-600 K then cooling in an applied electric field.  This
temperature range was chosen to avoid possible decomposition of the
sample while still reaching temperatures where the diffuse
scattering is nearly entirely suppressed.

     During thermal neutron measurements, two different experimental configurations were used
to investigate the structure and diffuse scattering both of which
employed a PG(002) monochromator fixed to reflect 14.7 meV
($\lambda$=2.36 \AA) neutrons. To optimize the momentum transfer
resolution $\delta q/Q$ of the structural measurements, the
horizontal beam collimations were set to (listed in order from the 
reactor wall to the detector and where $S$ refers to the sample) 10$'$-40$'$-$S$-20$'$-$open$
and the $d$-spacing of the (111) PMN Bragg peak was matched with
that of the (111) Bragg peak of a perfect SrTiO$_{3}$ analyzer for
which $\delta q/Q \sim 10^{-4}$.~\cite{Xu03:xx} During the diffuse
scattering measurements, the horizontal beam collimations were set to
40$'$-40$'$-$S$-40$'$-80$'$ and a PG(002) crystal was used as an
analyzer. This latter configuration provided significantly broader
resolution in momentum and was used only to study the diffuse
scattering and Bragg peak intensities.  A Pyrolytic Graphite filter
was placed before the monochromator to remove higher order neutrons.
Cold neutron measurements of the diffuse scattering were conducted
on the SPINS triple-axis spectrometer with collimations set to
$guide$-80$'$-$S$-80$'$-$open$.  The neutron final energy was set to
4.5 meV ($\lambda$=4.26 \AA) and cold Beryllium filters were placed
before and after the sample to remove higher order neutrons
reflected off the monochromator. The absorption of the neutron beam
was negligible in this experiment and therefore the neutron
diffraction measurements provided a means of investigating the bulk
properties of the PMN crystal.

    X-ray experiments were conducted at the X22A beamline at the
National Synchrotron Light Source, Brookhaven National Laboratory
using an incident photon energy of 10.7 keV.  A displex was used to
control the sample temperature.  The x-ray measurements were done in
a reflection geometry and we estimate the penetration depth of the
beam to be $\sim$ 15 $\mu m$.  The x-ray measurements therefore
provide a means of studying the properties of the near surface
region whereas neutrons probe the entire bulk crystal volume.  To
characterize the skin volume in more detail we conducted several
measurements using a 32 keV x-ray beam (with a penetration depth of
50 $\mu m$) which gave results consistent with those obtained using
a 10.7 keV beam as discussed below.  The x-ray results were
reproduced on a smaller crystal (0.5 cm $\times$ 0.5 cm $\times$0.1
cm) from the same boule as the larger crystal used for neutrons.
Measurements on this smaller sample gave the same results as that
measured on the larger sample, and will not be presented here.

\begin{figure}[t]
\includegraphics[width=7.5cm] {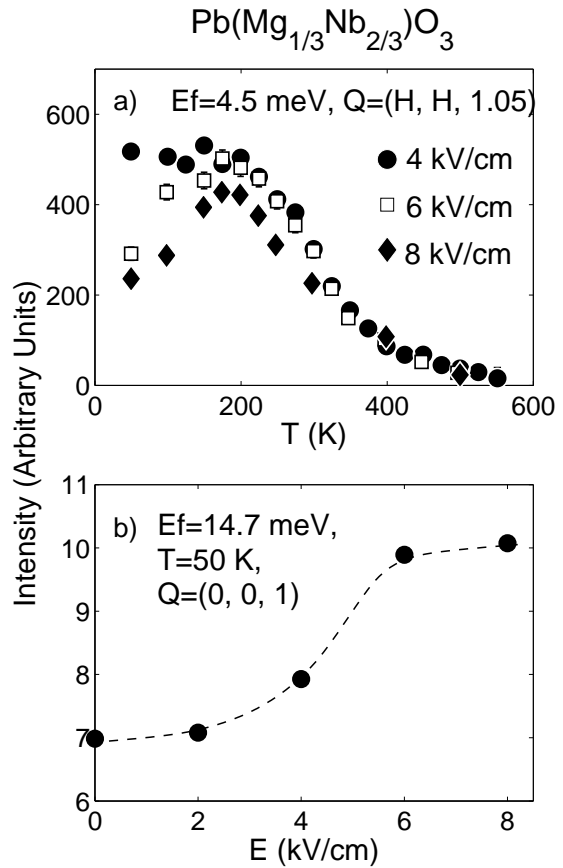}
\caption{\label{diffuse_bragg} $a)$ The temperature dependence of the diffuse
scattering is plotted for several field cooled values. Panel $b)$ illustrates the Bragg peak
intensity as a function of applied electric at 50 K.}
\end{figure}

\section{Diffuse Scattering}

    Two sets of diffuse scattering measurements were conducted on PMN under an
electric field.  Measurements on SPINS using cold neutrons focussed
on the temperature dependence and lineshape of the diffuse scattering near $\vec{Q}$=(001).  Measurements using thermal neutrons on
BT9 were conducted near $\vec{Q}$=(003) where the diffuse scattering
cross section is nearly an order magnitude larger allowing the
low-temperature lineshape to be investigated further in reciprocal space
from the Bragg peak. The possibility of
hysteresis and long time constants was investigated on the BT9
thermal triple-axis spectrometer near $\vec{Q}$=(003).

\subsection{Diffuse Scattering near $\vec{Q}$=(001)}

    To measure the temperature dependence of the diffuse scattering under
the application of an electric field we have studied the diffuse
scattering near the $\vec{Q}$=(001) Bragg peak with cold neutrons. Fig.
\ref{diffuse_001} shows contours of constant diffuse scattering intensity measured in
the (HHL) scattering plane near the (001) Bragg reflection and
representative linear scans through the two diffuse scattering rods.  In zero field, the diffuse scattering exhibits a symmetric butterfly-shape as previously observed with both x-rays and neutrons and
grows with cooling.~\cite{Hiraka04:70}  Under the application of an
electric field oriented along the [111] direction, a change in the
diffuse scattering is observed where the diffuse scattering perpendicular to the applied electric
field is preferentially suppressed.    We emphasize that the diffuse scattering does not
disappear, but it is substantially reduced relative to zero-field
cooled values.

    Figure \ref{diffuse_001} also shows scans through the diffuse streaks
around $\vec{Q}$=(001).  The two right hand panels show scans taken
after field cooling in a 6 kV/cm field.  The diffuse scattering is
initially suppressed after field cooling (upper panel), however, on removal of the
electric field (lower panel), the intensity returns to the value obtained under
zero field cooling conditions. Contrary to previous expectations
based on conventional ferroelectrics, this effect does not show any
electric field history dependence. Upon removal of the electric
field at low temperature, the diffuse intensity immediately returns to the
zero-field cooled values. This point will be discussed and quantified in
more detail in the next section, which deals with the strong diffuse
cross section near $\vec{Q}$=(003). Further measurements show that
zero-field cooling the sample to low temperatures and then applying
an electric field has the same affect as field cooling (in the same
field strength).

    In Fig. \ref{diffuse_bragg} we plot the temperature dependence of the
diffuse scattering under several strong electric fields. The data
shown are the intensities of linear scans along the (H,H,1.05)
direction.  The intensities were extracted by fitting each linear
scan to two symmetric Lorentzians displaced equally from the H=0 position.
Typical fits are shown in the right hand panels of Fig. \ref{diffuse_001}.
The diffuse scattering is onset
near the Burns temperature and appears to be insensitive to the
application of an electric field at high temperatures.  Below the critical temperature
T$_{c}$=210 K, the diffuse
scattering is strongly suppressed with increasing electric field,
while the (111) Bragg peak intensity increases (with field). The
suppression of the diffuse scattering accompanied by the enhancement
of the Bragg peak intensity suggests that the total scattering is
conserved and that the long-range (Bragg) and short-range (diffuse)
polar correlations are strongly coupled.  Therefore, as the
short-range polar order is suppressed (characterized by a reduction
in the diffuse scattering) an increase in long-range correlations
is observed (evidenced by the Bragg peak intensity).  This situation is
different from that proposed in PZN doped with 8\% PT where it was suggested
that a redistribution of the diffuse scattering occurs.~\cite{Xu06:74}

     These results demonstrate the presence of two distinct temperature scales.
At high temperatures above T$_{d}$ there is no diffuse scattering due to the lack of short-range polar correlations, while just below T$_{d}$ such diffuse scattering is unnaffected by
an electric field.  At lower temperature (below T$_{c}$), the application of an electric
field reduces the diffuse scattering and increases the Bragg peak intensity.  The existence of two characteristic temperature scales is an important observation
that is predicted by random field models of the relaxor ferroelectrics. Even
though some evidence for the two temperature scales has been
observed by investigating the phonons in PMN, a clear structural
signature has been lacking. An important question is whether all
the diffuse scattering is suppressed uniformly in reciprocal space, or one of the streaks or 
``wings" is preferentially suppressed as the data in the Fig.
lower left panel of \ref{diffuse_001} seem to suggest.  To answer these questions in
more detail, we study the diffuse scattering around $\vec{Q}$=(003).

\subsection{Diffuse Scattering near $\vec{Q}$=(003)}

\begin{figure}[t]
\includegraphics[width=8.75cm] {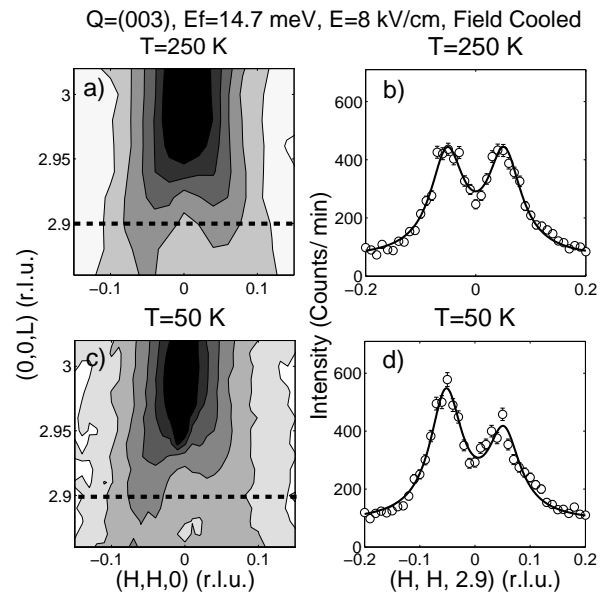}
\caption{\label{diffuse003_figure} The diffuse scattering near
$\vec{Q}$=(003) under the application of an 8 kV/cm electric field
along [111].  The left hand side shows diffuse contours plotted on a
logarithmic scale at temperatures above (250 K) and below (50 K)
T$_{c}$. The dark regions represent high intensity contours in comparison to lighter regions.  The right-hand panels display linear scans through the
diffuse scattering and are represented by the dotted line in the
contour plots.}
\end{figure}

\begin{figure}[t]
\includegraphics[width=9cm] {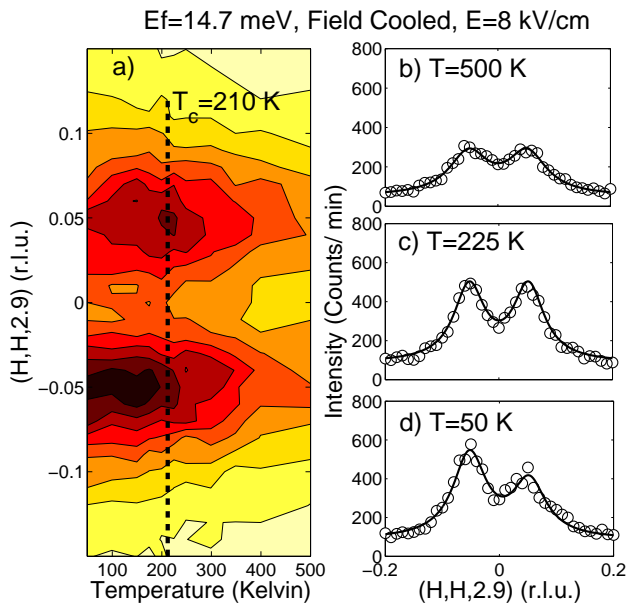}
\caption{\label{Q003_temp} (Color online) The temperature dependence of the diffuse scattering near $\vec{Q}$=(003).  Panel $a)$ is a contour plot (on a linear
scale) of scans along (H,H,2.9) measured in a field of 8 kV/cm as a function of temperature. The dark regions represent contours of high intensity compared with lighter regions.  Panels
$b)$ through $d)$ show examples of linear scans at 500 K, 225 K, and 50
K. The panels show the suppression of one wing of the diffuse
scattering at T$_{c}$.}
\end{figure}

\begin{figure}[t]
\includegraphics[width=9.25cm] {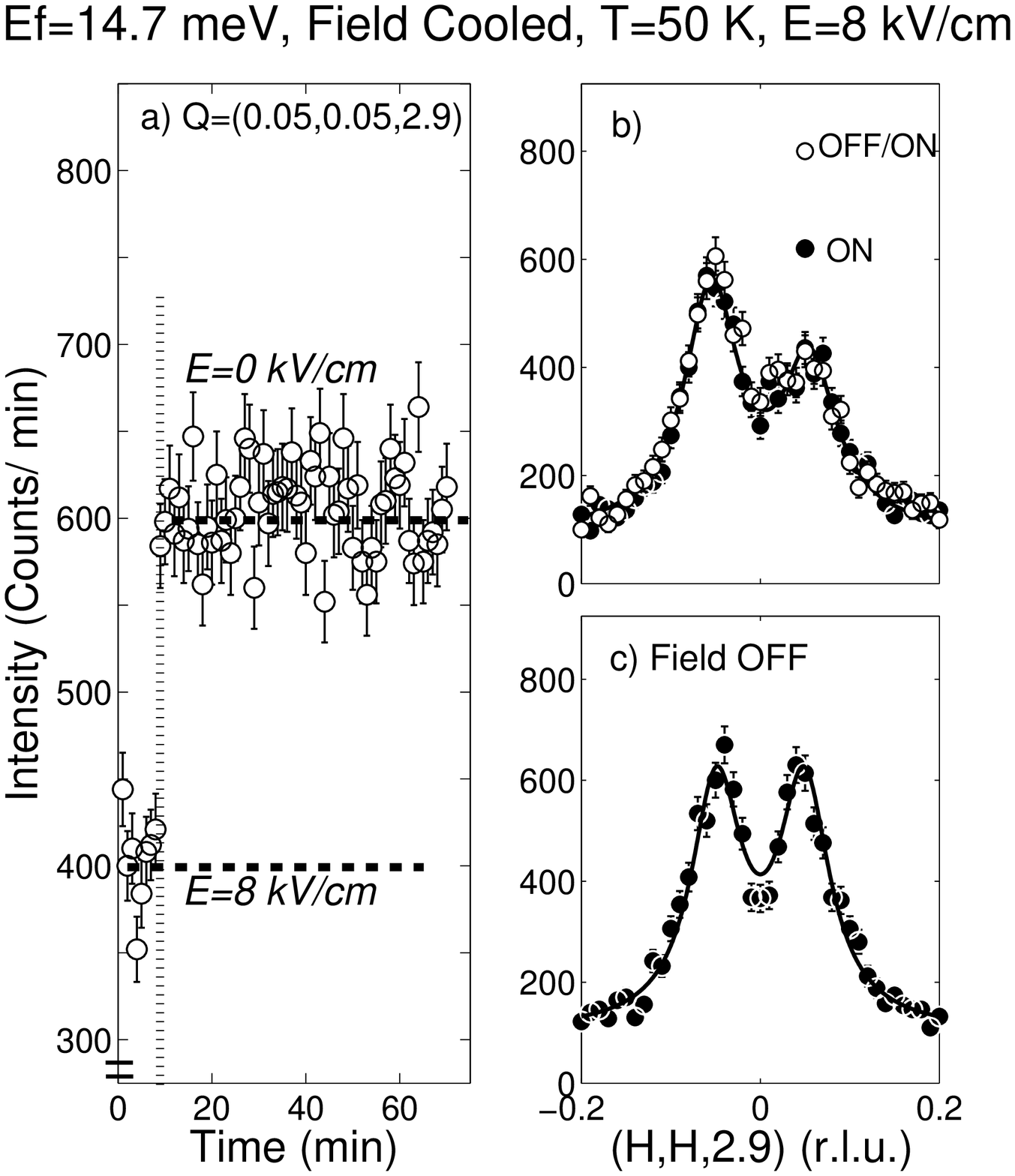}
\caption{\label{time_depend} Panel $a)$ plots the diffuse scattering
intensity at $\vec{Q}$=(0.05,0.05,2.9) as a function of time.  The
electric field was abruptly turned off and the leads electrically shorted to
ground at the time indicated by the dashed line. Panels $b)$ and
$c)$ show examples of linear scans of the diffuse scattering before and after
the field is turned off. The open circles in Panel $b)$ show the
diffuse scattering measured with the field turned back on $\sim$ 70
minutes after the field was turned off.  The diffuse scattering
reproduces the field cooled intensity.}
\end{figure}

    To analyze the electric field effect on the diffuse lineshape, i.e. the dependence on the momentum transfer $q$ (defined as the reduced momentum transfer measured relative to the Bragg peak) we have
studied the scattering near $\vec{Q}$=(003) using thermal
neutrons on the BT9 triple-axis spectrometer.  The $q$
dependence is plotted in Fig. \ref{diffuse003_figure} under field
cooled conditions (E=8 kV/cm) and at 250 K and 50 K (panels $a)$ and
$c)$ respectively). Data at large momentum transfers
(i.e. $L\sim 3$) were not obtained due to mechanical limits on the maximum scattering angle of the spectrometer. Examples of linear scans are represented on the
right hand panels $b)$ and $d)$. These scans were conducted both
above and below the critical temperature T$_{c}$=210 K.

    The data above T$_{c}$ at 250 K form the symmetric ``butterfly"
shape centered around $\vec{Q}$=(001) and are discussed above.  Below T$_{c}$
at 50 K, one wing of the butterfly is diminished as seen near (001)
in the lower left panel of Fig. \ref{diffuse_001}; however the larger diffuse
scattering structure factor near $\vec{Q}$=(003) provides a boost in intensity that allows the asymmetric lineshape to be
observed much more clearly because we can then measure the diffuse scattering further
away from the Bragg peak position. The asymmetric ``butterfly" shape is
clearly confirmed by the two linear $q$-scans measured along the (H,H,2.9) direction
illustrated in panels $b)$ and $d)$. At 250 K, two symmetric peaks
are observed, however at 50 K, one peak is suppressed.

    The temperature dependence of the field-cooled diffuse scattering near
$\vec{Q}$=(003) is illustrated in Fig. \ref{Q003_temp} as measured
by linear scans along the (H,H,2.9) direction.  Example linear scans
are represented in panels $b)$, $c)$, and $d)$.  The scans show
symmetric peaks displaced from the $H$=0 position above T$_{c}$=210
K and an asymmetric lineshape at low temperatures (below T$_{c}$)
representing the fact that one of the wings of the diffuse
scattering is diminished more than the other.  A contour plot
of the temperature dependence is illustrated in panel $a)$.  The contour plot shows the growth of
the diffuse scattering wings above T$_{c}$ as measured near
$\vec{Q}$=(001), however the plot shows that the asymmetry in the
diffuse lineshape is onset at T$_{c}$.  Therefore, at T$_{c}$ under
field cooled conditions, the diffuse scattering is reduced
preferentially along directions perpendicular to the applied
electric field.

    This feature was not obvious near $\vec{Q}$=(001) due to the fact
that the measurements there were conducted closer to the Bragg peak
and therefore less sensitive to changes in the lineshape. The increased
intensity of the diffuse scattering near (003) has allowed us to
investigate the diffuse scattering further away from the Bragg peaks
and therefore characterize the lineshape in more detail.  The
lineshape change can be understood in terms of domains and by noting
that the diffuse intensity is proportional to $(\vec{Q}\cdot \vec{\epsilon})^{2}$,
where $\vec{Q}$ is the momentum transfer and $\vec{\epsilon}$ is the
polarization vector. Domains with polarization antiparallel to the
applied field are not able to rotate and align with the field due to
the huge energy required for a 180$^{\circ}$ rotation. However,
domains for which the polarization is at a significantly smaller angle
(less than 90$^{\circ}$), require a much smaller energy to align
with the applied field.  This distinction is reflected in the data
at high electric fields.

    We have also investigated the possibility of the existence of a long
time scales and slow dynamics to the diffuse scattering near
$\vec{Q}$=(003) where the diffuse scattering is strong.  Long
time-scales have been observed in some examples of model magnetic systems
in random fields and in particular the case of
Fe$_{0.5}$Zn$_{0.5}$F$_{2}$ which is representative of a random
field Ising model.~\cite{Feng95:51,Hill91:66,Hill97:55} Long time scales were found to exist when the applied magnetic field (which in turn causes the presence of a random field as outlined by Fishman and Aharony in Ref. \onlinecite{Fishman79:12}) was turned on at low temperatures below the ordering transition.  It is not clear in our case if the same situation would apply as the random fields are presumably introduced through structural disorder and hence present at all temperatures and at all applied electric fields.  Long time
dependences have been measured in the birefringence and structural properties (characterized with low-energy x-rays) in PMN (Ref.
\onlinecite{Westphal92:68,Vak97:103,Calvarin95:165}), in the relaxor SBN (Ref. \onlinecite{Chao05:72}), and KTaO$_{3}$ doped with 3\%\ Li (\onlinecite{Yokota:07:19}). Our data for PMN is
illustrated in Fig. \ref{time_depend}. Panel $a)$ plots the time
dependence of the intensity at $\vec{Q}$=(0.05,0.05,2.9) which (as can be
seen from Fig. \ref{diffuse003_figure}), corresponds to the peak of one of
the diffuse scattering wings.  The temperature in Fig.
\ref{time_depend} is 50 K and the sample was cooled in an 8 kV/cm
from high temperature.  At time $t=0$ the sample is in a field cooled
state and then the field was turned off and the surfaces of the crystal
electrically grounded at the time indicated by the dotted line.  The diffuse
scattering intensity rose abruptly and did not change for another
hour of measurements. Linear scans showing the diffuse scattering
intensity when the electric field was on (field cooled and at a time
before that indicated by the dotted line in Fig. \ref{time_depend})
and then off (after waiting for about 1 hour after the field was
turned off and crystal faces electrically shorted to ground) are represented in
panels $b)$ and $c)$ respectively by the filled circles.  After
turning the electric field off for over an hour, we then turned it
back on at 50 K and the resulting profile is represented by the open
circles in panel $b)$.  The diffuse intensity tracks the original
intensity obtained under field cooling conditions within error.  We
note that the Bragg peak intensity responded similarly but with an
increase in intensity on application of the electric field, and
subsequent decrease on removal (as expected given our measurements
made near $\vec{Q}$=(001)).  We therefore conclude from this
sequence, that there are no long time scales associated with the
electric field effects measured here.

    The lack of any time dependence contrasts with that found using
birefringence and x-rays in PMN (Ref. \onlinecite{Westphal92:68,Vak97:103,Calvarin95:165}) and also
neutron diffraction measurements on PZN doped with PT (Ref.
\onlinecite{Xu05:72}).  Doping with PT has been snown to stabilize
the rhombohedral phase in PMN and related materials; therefore the rhombohedral phase
and the resulting effects on the diffuse scattering maybe more
easily frozen in for those systems.  The birefringence measurements
are more difficult to reconcile, however, we emphasize that with
neutrons we are measuring at a non zero wave vector associated with a
certain length scale. The length scales studied with neutrons and
the expected size of the polar nanoregions have been discussed by
Vugmeister.~\cite{Vug06:73}  In the case of the linear scans measured
along (H,H,2.9) we would expect to study polar correlations with a
length scale $\lambda \sim 1/q \sim 6$ \AA, which is a very short length
scale compared to other techniques.  It therefore is possible that
the time dependence measured with birefringence is associated with
longer length scales than investigated here. Indeed, studies on the
history dependence of PMN doped with PT have found domain formation
and history dependence associated with length scales as large as
millimeters.~\cite{Bai05:97,Viehland04:96}  Our neutron diffraction
experiments are not sensitive to such long length
scales.  These conflicting results may also indicate the presence of
two different behaviors for the bulk and surface.  Neutrons probe
the bulk where optical techniques are confined to a surface skin
region.  This point is investigated in the next section (and is compared directly with x-ray measurements similar to those conducted in Refs. \onlinecite{Vak97:103,Calvarin95:165}), which discusses
the shape of the unit cell and structural properties.

    The absence of any history dependence suggests that the ground state
of PMN under field cooling is not a glassy state nor in a well defined ferroelectric ground state.~\cite{Zhao07:75}  Here long-range and
short-range polar order coexists and compete with each other.
Similar measurements on single crystals of PZN and PZN-8\%PT do not
show a suppression of diffuse scattering intensities, but only a
redistribution with field history dependence. One distinct
difference between PMN and PZN is that in the latter, long range
rhombohedral domains form with an external [111] field. Therefore
the memory effect of diffuse scattering measured in PZN and
PZN-8\%PT is more likely to be associated with ferroelectric domain
formation and rotation.

    The electric field effects show that there is a direct relation between
the polar correlations and the diffuse scattering.  The fact that the diffuse
scattering is only diminished below T$_{c}$ and, in particular, very anisotropic
(being elongated strongly along one direction) and exhibits no memory is
suggestive that the diffuse scattering is directly related to the polar
structure in the relaxor ferroelectrics as described in several
models.~\cite{Welberry05:38,Xu04:69}

    An important result is that our measurements show a \textit{clear
structural signature} of two distinct temperature scales. At a high
temperature, the diffuse scattering is onset and is insensitive to
the application of an electric field.  Presumably, in this
temperature range the polar moments are isotropic and insensitive to
the presence of a strong electric field which selects a particular
direction.  This suggests that the polar moments have a continuous
symmetry between the temperature range T$_{d}$ and T$_{c}$.  In
analogy to spins in magnetic systems, the Hamiltonian in this
temperature range can be approximated by Heisenberg component.
Below T$_{c}$, the situation and the underlying Hamiltonian change
as evidenced by both a decrease in the diffuse scattering intensity
and the asymmetric lineshape measured near $\vec{Q}$=(001) and
(003). Both the decreases in intensity and asymmetry are onset at
T$_{c}$. This suggests a change in the universality class as the
polar moments have a strong anisotropy implying that a discrete
symmetry is dominating the ground state Hamiltonian, as is the
case for a cubic anisotropy term in the Hamiltonian.  Given that the
same Hamiltonian must apply at all temperatures, these results show
that T$_{c}$ characterizes the cubic anisotropy in the Hamiltonian,
while T$_{d}$ describes the continuous (Heisenberg) component of
the Hamiltonian. Therefore, PMN is a system where the energy scale
of the continuous component of the Hamiltonian dominates over the
anisotropic component.  The dominant term (and hence the effective
universality class) can be tuned with temperature.

    PMN (and indeed all relaxors) exhibit a significant amount of
structural disorder as a result of the mixture of two different ions
with different valency.  In the case of PMN, we expect there to be a
random field introduced as a result of the mixture between Mg$^{2+}$
and Nb$^{5+}$ ions.~\cite{Kleeman06:41}  Random fields have very different effects on
universality classes with discrete and continuous symmetries as
summarized in the work by Imry and Ma.~\cite{Imry75:35} For the case
of discrete symmetry (for example in PMN below T$_{c}$ where a
strong cubic anisotropy dominates), the system should be less
sensitive to the formation of random fields and long-range polar
correlations are predicted to develop.  In the case of a continuous
symmetry (as in PMN above T$_{c}$) the correlations should be
strongly affected and short ranged in the presence of a random
field.  These two cases are reflected in our data with long-range
polar correlations only developing below T$_{c}$ as evidenced by an enhancement of the Bragg peak intensity.

    Random field models of relaxors predict two distinct temperature scales
associated with a high-temperature transition and then a
low-temperature cross over point where the universality class
changes from being Heisenberg to Ising like in analogy with model
magnetic systems.~\cite{Stock05:74,Stock04:69} While it was
suggested that the phonon spectrum may well reflect the two
temperature scales required by such a model, the structural
properties remain ambiguous as the diffuse scattering from the bulk (measured with neutrons) in PMN shows little sign of T$_{c}$ and PZN is heavily contaminated by a large
surface effect.   We note that low-energy x-ray measurements probing the surface of PMN have found evidence for an anomaly of the diffuse scattering intensity near T$_{c}$.~\cite{Dkhil01:65}  No such clear anomaly is observed in our neutron data.  The two temperature scales required by random
field models are reflected in the diffuse scattering intensity as a
function of temperature under the application of a strong electric
field.  At high temperatures where the universality class is
Heisenberg, random fields destroy the phase transition and
short-range correlations are formed.  It is only at T$_{c}$, where
the system crosses over to an Ising universality class, that the
phase transition becomes robust to random fields and long-range
correlations can develop.  We note that this model still requires further verification as some measurements have found evidence for more than two temperature scales in PMN.~\cite{Dkhil01:65}  It maybe that some of these temperature scale can be reconciled based on the dynamics and the energy resolution of the experimental probe, however further experimental work is required to investigate this point.

\section{Structural Properties}

\begin{figure}[t]
\includegraphics[width=8.0cm] {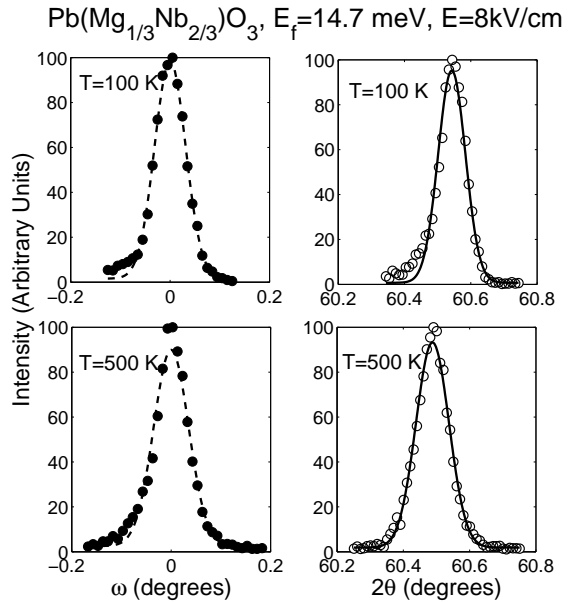}
\caption{\label{srtio3_figure} Longitudinal ($\theta-2\theta$) and
transverse ($\omega$) scans through the (111) Bragg peak at 100 K
and 500 K are displayed. No strong distortion at low temperatures is
observed.}
\end{figure}

    Two sets of experiments were conducted under field cooling conditions to search for a structural
distortion under an electric field in field cooling conditions.  To
study the bulk properties of the PMN crystal, we have used a
high-resolution neutron diffraction setup.  For PMN, neutrons have a
large penetration depth on the order of $\sim$ 1 cm, therefore
providing a bulk probe of the structure.  These measurements found
no evidence for a structural distortion with a sensitivity estimated
at 90$^{\circ}$-$\alpha$= 0.03$^{\circ}$.

    The second set of measurements were conducted using 10.7 keV x-rays
in reflection geometry with a penetration depth of $\sim$ 15 $\mu
m$.~\cite{penetrate} These measurements were aimed at studying the
surface properties of the crystal.  Evidence for a structural
distortion was obtained with 90$^{\circ}$-$\alpha$ measured to be
0.08 $\pm$ 0.03$^{\circ}$. We emphasize that these experiments were
conducted under field-cooled conditions where the sample was heated
to temperatures between 550-600 K (above the onset temperature of
the diffuse scattering and the polar correlations) and then cooled
under the application of an electric field.  This field cooling
sequence is very different from that used in previous studies (for example Ref.
\onlinecite{Vak97:103}) where the field is applied at room
temperature or below the onset temperature of diffuse
scattering.

\subsection{Unit cell in the bulk (Neutron diffraction E$_{f}$=14.7 meV and SrTiO$_{3}$ analyzer)}

    The field effect on the diffuse scattering is strongly correlated with
T$_{c}\sim 210$~K and not the Burns temperature, showing directly
that T$_{c}$ is an important temperature scale below which
long-ranged polar correlations are enhanced by the external field.
On the other hand, the true origin of the long-range order is
controversial. Fig. \ref{srtio3_figure} shows longitudinal scans
through the (111) Bragg peak in PMN, measured after field cooling
from 600~K. For all electric field strengths up to $E=8$~kV/cm, we
observe no change in the radial lineshape near (111), as opposed to
the expected splitting due to a rhombohedral distortion. We have
checked other equivalent \{111\} reflections and have always found a
single peak with the same $2\theta$ value (within an error $\pm$
0.01$^{\circ}$ based on Gaussian fits) and hence the same
$d$-spacing, ruling out the possibility of a rhombohedral single
domain state.  We note that all of these measurements were conducted
under the field cooling conditions outlined above and low
temperature measurements were made after waiting more than 2 hours
at low temperatures.  This rules out the possibility of a long time
dependence and slow response as suggested in some x-ray diffraction
studies.~\cite{Vak97:103}  This provides definitive evidence that
the unit cell shape in pure PMN with field cooling (up to 8kV/cm) to
low temperature is still cubic on average, or, much (more than an
order of magnitude) less rhombohedral than previously
reported~\cite{Ye98:81}.

\begin{figure}[t]
\includegraphics[width=8.0cm] {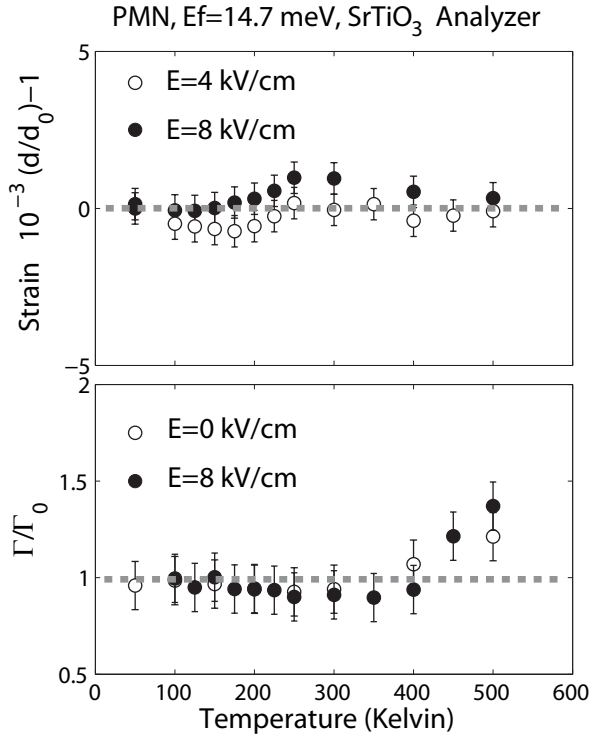}
\caption{\label{cte_bt9} The lattice strain is plotted in the upper
panel as a function of temperature for field cooling with
electric fields of 4 and 8 kV/cm.  The values for $d_{0}$ are taken as the
measured lattice constant in zero field cooled conditions at the given
temperature.  The longitudinal width as a
function of temperature for $\vec{Q}$=(111) Bragg peak is plotted in
the lower panel $b)$. No change in the longitudinal linewidth or anomalous
strain is observed in the bulk under field cooled conditions.}
\end{figure}

    Further confirming the lack of any clear rhombohedral distortion we
plot the strain (upper panel) and the longitudinal width measured at
the (111) Bragg peak (lower panel) as a function of temperature in
Fig. \ref{cte_bt9}. From the temperature dependence, it can be seen
that there is no change of the thermal expansion coefficient on the
application of an electric field (as discussed in terms of the
expected sensitivity of the experiment below).  Also, for large
fields (8 kV/cm) the strain with temperature is no different to that
measured at lower fields. The upper panel shows the lattice strain
as a function of temperature taking $d_{0}$ to be the unit cell
length measured at the given temperature in zero field cooling
conditions. The lower panel illustrates the linewidth as a function
of temperature. The values are normalized to $\Gamma_{0}$ which is
taken as the average value over the entire temperature range
studied.  Both the lattice strain and the longitudinal width show no
strong response to an electric field nor T$_{c}$.   Therefore, we do
not observe a clear structural transition to a long-ranged ordered
rhombohedral phase at T$_{c}$.

    It is important to establish an upper limit on the sensitivity of the
neutron measurements to a rhombohedral distortion and define this
quantitatively.  One way of characterizing the sensitivity is based
on the half-width at half maximum in Fig. \ref{srtio3_figure}.  If
the crystal did undergo a structural distortion, we might expect the
Bragg peak to split in the longitudinal direction as measured in a
$\theta$-2$\theta$ scan at $\vec{Q}$=(111).  This splitting would be
result of domains forming with different $d$-spacings for the (111)
and ($\overline{1}11)$  Bragg peaks.  Since we only observe one peak
in any such scan at low temperatures, the half-width at half maximum
provides a good measure of the sensitivity to such a scenario.
Based on this we put an upper limit of 0.02$^{\circ}$ for the value
of the crystallographic parameter $90^{\circ}-\alpha$.

    Another method of defining the sensitivity of the experiment is to
characterize how sensitive the results are to the formation of a
single domain state.  Such a ground state is not unexpected given
the large field strengths applied here.  In this case, we would
expect an anomaly in the thermal expansion at low temperatures with
different domains having different $d$-spacings (as discussed
above). In the upper panel of Fig. \ref{cte_bt9}, the strain is
plotted as a function of temperature for field cooling under
electric fields of 4 kV/cm and 8 kV/cm. Based on the standard
deviation of the data (from the case of no strain which is
$d/d_{0}-1$=0) we estimate that the we are sensitive to values of
90$^{\circ}$-$\alpha$ $>$ 0.03$^{\circ}$.

    We have therefore defined the sensitivity of the experiment to the formation of a ground state
with domains, and a single domain rhombohedral domain.  Based on
these two methods for defining the sensitivity of the experiment to
a rhombohedral distortion, we therefore expect that the high
resolution neutron diffraction measurements are sensitive to
distortions greater than 90$^{\circ}$-$\alpha$ $\sim$
0.02-0.03$^{\circ}$. We note that PMN doped with 27\% PT undergoes a
transition to a rhombohedral phase with $\alpha \sim$0.18$^{\circ}$ and in the
surface of pure PZN, a distortion of
90$^{\circ}$-$\alpha$=0.085$^{\circ}$ is measured.~\cite{Xu03:68} PMN
doped with 10\%PT has been measured to undergo a distortion with
90$^{\circ}$-$\alpha$=0.13$^{\circ}$.~\cite{Dkhil01:65}  In pure PMN, x-ray
measurements have found
90$^{\circ}$-$\alpha$=0.1$^{\circ}$.~\cite{Ye98:81} In all cases, we
expect that the current experiment would have been able to detect
such distortions based on our estimates for the sensitivity.

\begin{figure}[t]
\includegraphics[width=9.0cm] {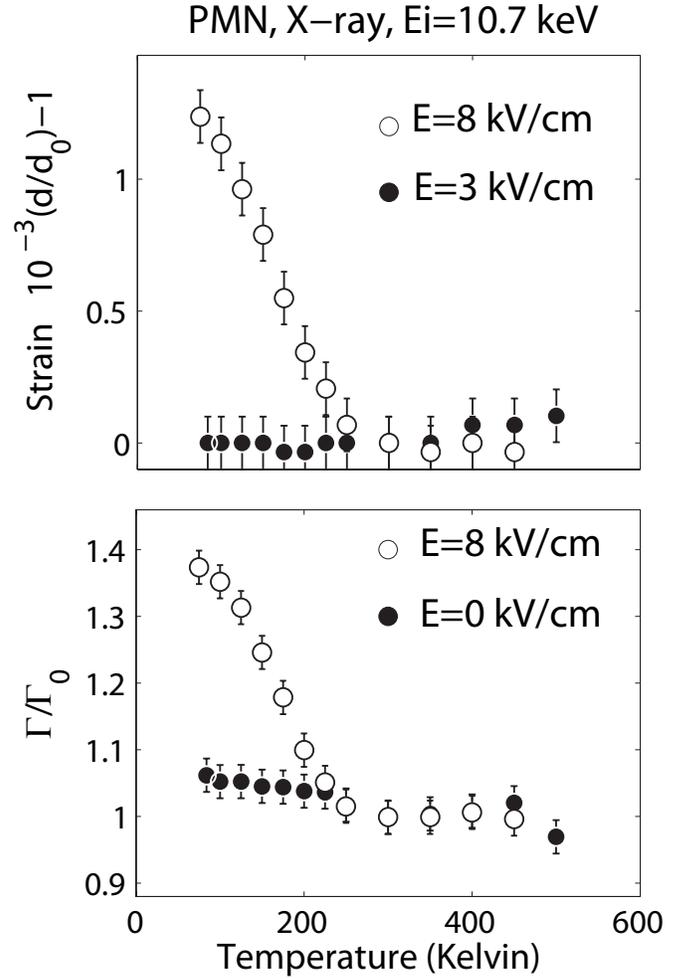}
\caption{\label{summary_10keV} The lattice strain is plotted in the upper panel
as a function of temperature for field cooling with electric fields of 3 and 8
kV/cm.  The values for d$_{0}$ are taken as the lattice constants measured under
zero field cooling conditions at the given temperature.  The lower panel plots
the linewidth as a function of temperature for cooling under electric fields of
0 and 8 kV/cm.  The data shows a clear anomaly in both the lattice strain and
the linewidth at T$_{c}$ for cooling under a 8 kV/cm field.}
\end{figure}

\subsection{Unit cell in the skin (X-ray diffraction with Ei=10.7 keV and in reflection geometry)}

    To reconcile our results with previous
measurements~\cite{Calvarin95:165,Ye98:81}, where a rhombohedral
distortion was observed in PMN with field cooling, we consider work
done on PMN and PZN.  It was shown that in both systems there is a
``skin'' region on the order of tens of microns thick, having a unit
cell shape different from that of the
bulk.~\cite{Conlon04:70,Xu06:79,Gehring04:16}  In zero field, PZN
undergoes a structural distortion in the near surface region only.~\cite{Lebon02:14}
 The surface thickness was estimated in these experiments based on
the penetration depth of the x-rays used and also using a small
narrow neutron beam in a strain scanning geometry.

    In Fig. \ref{summary_10keV}, we investigate the possibility that
only the near surface region of PMN undergoes a structural
distortion in an electric field.   To do this we use low-energy x rays in reflection geometry to study the surface of PMN under an electric field.  Fig. \ref{summary_10keV} shows
the strain (upper panel) and longitudinal linewidth (lower panel) of
the (111) Bragg peak as a function of temperature for various
electric fields. The value of $d_{0}$ was taken to be the lattice
constant measured at the given temperature under zero field cooling
conditions.  The linewidth was extracted by fitting the profiles to
a Lorentzian raised to the power of $3/2$. This lineshape is
expected for the case of random fields introduced by domains with
the spectrometer integrating the scattering perpendicular to the
scattering plane. Here low-energy 10.7 keV x-rays do show a
broadening of the Bragg peaks at low-temperatures under field cooled
conditions. The results also show the lattice strain grows below
T$_{c}$ under large electric fields, in contrast to neutrons which
show no anomaly in the strain around T$_{c}$.  This is indicative of
a structural transition into a more rhombohedrally distorted phase
at T$_c$.  Because of the presence of one symmetric peak displaced
in 2$\theta$, we interpret this in terms of a single domain state.

\begin{figure}[t]
\includegraphics[width=9.0cm] {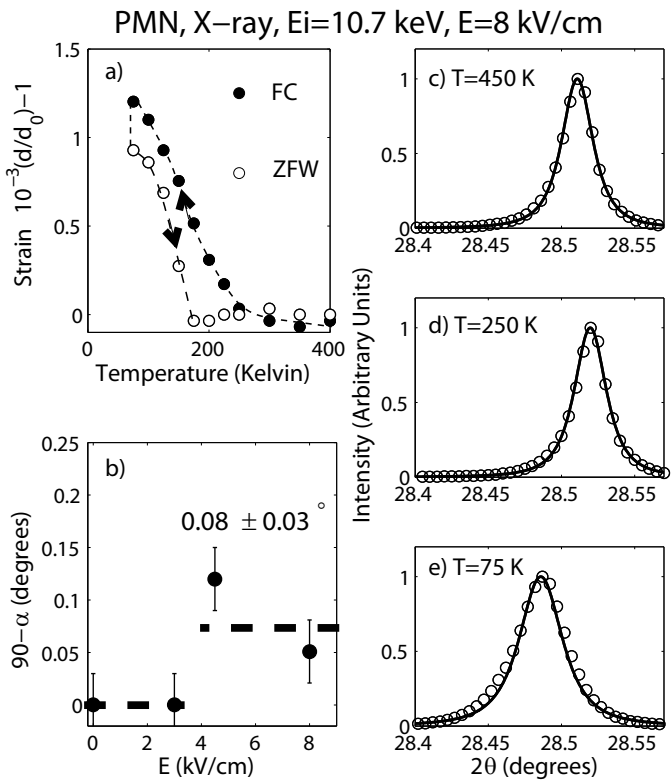}
\caption{\label{xray_distort} Panel $a)$ plots the lattice strain
under field cooling (8 kV/cm) and zero field warming conditions. The
data shows that the distortion remains on removal of the electric
field at low temperatures.  Panel $b)$ plots the measured value of
$\alpha$ as a function of electric field. Panels $c)$ through $e)$
shows $\theta - 2\theta$ scans through the (111) Bragg peak at 450
K, 250 K, and 75 K. }
\end{figure}

    Data illustrating the structural distortion in more detail are
illustrated in Fig. \ref{xray_distort}.  Panel $a)$ illustrates the
lattice strain under field cooling in an 8 kV/cm field, and warming
after removal of the electric field at low temperatures.  Based on
this measurement it can be seen that on removal of the field at low
temperatures, the lattice strain remains but on warming a hysteresis
is observed in the response.  This effect is expected and has been
measured previously in PMN where a rhombohedral distortion was
observed at low temperatures.\cite{Calvarin95:165,Ye98:81} Such a memory
effect was not observed in the bulk with neutrons and further
distinguishes the surface skin layer from the bulk.   We also note
that on removal of the electric field at low temperatures and
heating, the Bragg peaks remained broad and only recovered on
warming through T$_{c}$. Therefore, the surface region displays the
expected memory effect which is absent in the bulk and is similar to
previous results using x-rays.

    Panel $b)$ of Fig. \ref{xray_distort} shows the value for $\alpha$ as a function of applied
electric field and shows that at large values of the electric field
$\alpha$=0.08$\pm$0.03$^{\circ}$. Such a large distortion is well
within the sensitivity of our neutron diffraction measurements
further emphasizing that the surface undergoes a much stronger
distortion than the bulk phase.  The values of $\alpha$ are similar
to those measured previously in thin crystals of PMN and the
temperature at which the distortion is observed, is also consistent
with previous data.  We note that no such transition was observed in
the bulk single crystals with neutrons.

    Panels $c)$ through $e)$ of Fig. \ref{xray_distort} illustrate raw $\theta$-2$\theta$ scans
through the (111) Bragg peak using 10.7 keV x-rays.  Panels $c)$ and
$d)$ are taken at temperatures above T$_{c}$=210 K and show the
expected decrease of the lattice constant on cooling.  However,
panel $e)$ shows a scan at low temperatures below T$_{c}$ and shows
that the lattice as expanded.  The strain measurements presented in
Fig. \ref{summary_10keV} show that this occurs at T$_{c}$ and is
suggestive of the formation of a single domain rhombohedral ground
state.

    The x-ray results taken in comparison with the neutron data show
that the ferroelectric distortion is confined to the near surface
region of the sample.  In an effort to determine how thick this skin
layer is, we have repeated the experiment with 32 keV x-rays on the
same beamline X22A (NSLS).  The penetration depth in this case is
$\sim$ 50 $\mu m$.  These results also showed a structural distortion
below T$_{c}$ with 90$^{\circ}$-$\alpha$=0.06$\pm$0.03$^{\circ}$.  This
distortion is within error equal to that obtained with the lower
energy x-rays of 10.7 keV.  Therefore, the skin layer in PMN is at
least 50 $\mu m$ deep, however our neutron results show it is not
characteristic of the bulk.

    These results supports the claim that the distortion is limited to a small
region near the surface and is not representative of the bulk
properties and indicate that the near surface region is sensitive to
the application of an electric field.  This is in accord with
previous x-ray measurements (with a similar penetration depth to the
experiments presented here) on PMN which were conducted on 30 $\mu
m$ thick samples and comparable to the estimated thickness of the
``skin'' layer in our sample.~\cite{Ye93:145,Conlon04:70} Models based on different dual structures for the PZN-$x$pT phase diagram have been proposed and discussed.~\cite{Shvartsman05:71}  Our
results are strongly suggestive that anomalies in dielectric
measurements may also be dominated by a skin region and not an
indicative of the bulk phase. These results are also consistent with 
recent piezoresponse force microscopy which suggest unusual 
properties near the surface of PMN-10\% PT.~\cite{Shvartsman07:101}  

    Even though the presence of a surface skin having different
critical properties is unusual in materials and condensed matter
physics, the effect has been observed before in the case of model
magnets in the presence of random fields.  The experiments described
here are reminiscent of the situation in the three dimensional
random field Ising model applied to Mn$_{0.75}$Zn$_{0.25}$F$_{2}$. A
series of experiments by Hill $\textit{et al.}$ (Ref.
\onlinecite{Hill91:66}) showed the presence of long-range magnetic
order in a skin (measured through magnetic x-ray diffraction) in
contrast to the bulk which only showed short-ranged magnetic order,
exactly the same situation here.  Also, there have been many systems
which exhibit the two-length scale effect and have a skin which
displays different critical behavior than the bulk. Examples include
Tb (Ref.\onlinecite{Hirota94:49,Gehring93:71}), SrTiO$_{3}$ (Ref.
\onlinecite{Cowley78:11}), and CuGeO$_{3}$ (Ref. \onlinecite{Wang01:63}). Possible explanations of the two-length
scales have been proposed and a review and discussion can be found
elsewhere.~\cite{Cowley96:66}

    It is interesting to speculate whether
the presence of a skin is important (and even necessary) for the
presence of a relaxor ground state. Given the many results showing
the presence of a distinct skin in relaxors, it maybe that the
ferroelectric transition in these systems is driven by the surface
boundary condition.  Whether this is the result of strain or
chemical inhomogeneity at the surface is not clear and further
investigation would be required to distinguish these two cases.

    It is interesting to note that even when the polar nanoregions are
reduced by the field, a long-range (rhombohedral) ferroelectric
phase has not been established in the bulk. However, as noted in the
section discussing the diffuse scattering, the Bragg peak intensity
is enhanced at low temperatures as a function of field implying a
more ordered structure.  This occurs while the diffuse scattering is
reduced under an electric field. We note that the absence of a long
range ferroelectric ground state is supported by recent NMR results
which show at least half of the sample remains in a disordered
ground state under the application of an electric field as discussed
in the introduction.

\section{Conclusions}

    We have presented a neutron and x-ray scattering study of the diffuse
scattering in PMN under the application of a strong electric field.  Our results
show that the diffuse scattering is directly associated with short-range polar
order in the system. The balance between this short-range polar order and long-
range atomic order in the bulk of the sample is reflected in a trade off of
intensity between the diffuse and Bragg scattering and is tuned with an electric
field.  However, for fields below 8 kV/cm, no long-ranged ferroelectric phase is
formed.  The absence of any hysteresis in the bulk suggests that the ground
state of PMN is not a glassy state. Instead, the presence of two temperature
scales agrees with predictions of recent random field models proposed for
relaxor systems.  The results also show that the outer-most tens of microns
``skin'' layer behaves distinctly different  than the bulk (under field).

\begin{acknowledgements}

We would like to thank S.M. Shapiro, R. Cowley, J. Hill, and I. Swainson for
helpful discussions. We thank B. Clow and B. Schoenig for important
technical help in operating the electric field setups. We also thank
J. Thomas for assistance during the x-ray experiments on X22A. We
acknowledge financial support from the Natural Science and
Engineering Research Council of Canada and through DMR-9986442 and
from the U.S. DOE under contract No. DE-AC02-98CH10886, and the
Office of Naval Research under Grant No. N00014-99-1-0738.

\end{acknowledgements}


\end{document}